\documentclass[fleqn,twoside]{article}
\usepackage{espcrc22}

\usepackage{graphicx}
\newcommand{\AmS}{{\protect\the\textfont2
  A\kern-.1667em\lower.5ex\hbox{M}\kern-.125emS}}

\title{Catching a bullet: direct evidence for the existence of dark matter}

\author{D. Clowe\address[OU]{Department of Physics and Astronomy, Ohio University, 251B Clippinger 
Lab, Athens, OH 45701, USA},
        S. W. Randall\address[CFA]{Harvard-Smithsonian Center for Astrophysics, 60 Garden Street, 
Cambridge, MA 02138, USA}
        and
        M. Markevitch\addressmark[CFA]}

\begin{document}

\begin{abstract}
We present X-ray and weak lensing observations of the merging cluster system 1E0657$-$556.  Due to the recently collision of a merging subcluster with the main cluster, the X-ray plasma has been displaced from the cluster galaxies in both components.  The weak lensing data shows that the lensing surface potential is in spatial agreement with the galaxies ($\sim 10\%$ of the observed baryons) and not with the X-ray plasma ($\sim 90\%$ of the observed baryons).  We argue that this shows that regardless of the form of the gravitational force law at these large distances and low accelerations, these observations require that the majority of the mass of the system be some form of unseen matter.
\vspace{1pc}
\end{abstract}

\maketitle

\section{Introduction}

\begin{figure*}[t]
\includegraphics*[width=6in]{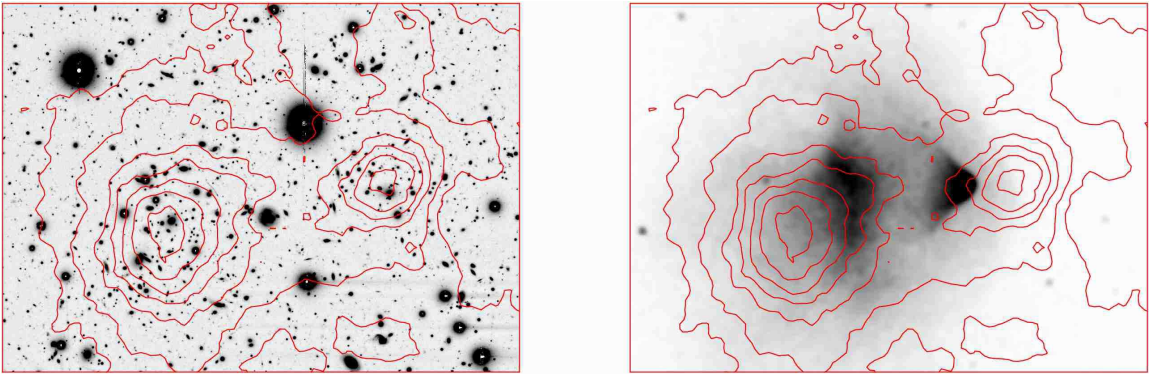}
\caption{{\it Left panel:} $R$ band image from Magellan of the merging cluster 1E0657$-$558.  {\it Right panel:} 522 ks {\it Chandra} image.  Shown in contours on both panels is the weak lensing $\kappa$ reconstruction.}
\end{figure*}

We have known since 1933\cite{ZW33.1} that clusters of galaxies have gravitational potentials which are
too large to be explained by the mass of visible baryons under the assumption that gravity obeys the same $1/r^2$ force law as it does locally in the inner solar system.   In most situations, however, all the matter which could be causing the gravity in a given cluster occupies the same volume, so there is a degeneracy between whether the gravity comes from some type of matter does not emit, absorb, or deflect light in any observed bandpass (''dark matter''), or comes from only the observed baryonic mass and the excess is due to the gravitational laws being different at large distances and/or small acceleration levels.

The dark matter paradigm has generally been the preferred solution in cosmological models, mostly due to alternative gravity models historically being ad hoc explanations for certain sets of observations on specific types of objects, such as spiral galaxy rotation curves, and it was difficult to extend these models to other observations.  In the past few years, however, several alternative gravity theories have been developed which are derived from modifications to general relativity, and can be used to model different observations\cite{BE04.1,BR06.1}.  Initial studies suggest that these theories can reproduce at least the gross properties of many extragalactic and cosmological observations.  One is left with either comparing how well the various theories do at explaining the fine details of the observations or finding objects in which the dark matter, if it exists, would be physically separated from the visible matter, and therefore could be detected directly by it's gravitational potential.  One type of object where this separation occurs in is merging galaxy clusters.

In clusters of galaxies, $\sim 90\%$ of the visible baryons are in a million degree plasma which fills the cluster volume.  The other $\sim 10\%$ are in the stars in the cluster galaxies.   During a collision of two clusters, the galaxies are effectively collisionless particles affected only by tidal forces while the plasma clouds are highly collisional and therefore are slowed by ram pressure.  Thus, after the collision the galaxies will be leading the now slower moving plasma.  If dark matter particles are also collisionless, as is widely assumed, any dark matter present in the system would be located near the galaxies.  Therefore, if one could measure a spatially resolved  gravitational potential of such a system, it could be directly determined whether and how much dark matter exists.

One such merging cluster system is 1E0657$-$558 ($z = 0.296$).  In this system, a smaller merging cluster has fallen through a larger cluster and is currently moving away.  The core of the plasma cloud of the merging subcluster, ''the bullet,'' has survived the passage and has a Mach 3 bow shock in front of it \cite{MA02.1}, from which we have derived that the bullet is currently moving at $\sim 4700$ km/s  relative to the main cluster's plasma cloud\cite{MA06.1}.  Because the line-of-sight velocity different of galaxies in the structures is only $\sim 600$ km/s \cite{BA02.1}, we know that this merger is occurring within a few degrees of the plane of the sky.    This system therefore provides an excellent opportunity to test if we can directly detect the presence of dark matter from it's gravitational signature via weak gravitational lensing.

\subsection{X-ray Observations}

The cluster was observed with the {\em Chandra}\/ X-ray                                                
observatory in 2002--2004 for an exceptionally long total                                              
exposure of 522 ksec (useful time). The ACIS-I $16'\times                                              
16'$ field of view covers the entire cluster.                                        
Full technical details of the X-ray analysis will be given                                             
in elsewhere\cite{MA06.1}.  Given the gas                                                  
temperature between $T=7-30$ keV across the cluster\cite{MA02.1,MA06.1} and the peak ACIS-I                                                    
sensitivity at $E=1-2$ keV, the 0.8--4 keV band image used                                             
in this work essentially gives the line of sight                                             
integral of the square of the gas density, with only a small                                           
correction due to the gas temperature differences.                                                     
                                                                                                       
The gas mass distribution was derived by fitting a 3D                                           
geometric model to the X-ray image, consisting of a                                                    
component for every physically distinct region.  The                                           
model components are assumed to be cylindrically symmetric in the plane                                            
of the sky.    The                                           
gas bullet was modeled with a shuttlecock morphology that is                                           
densest near the tip and falls off with distance from the                                              
tip and distance from the surface of the shuttlecock. The shock front is fit with a                                                  
similar model, and is truncated at the interface with the                                              
bullet cold front. The eastern brightness peak, elongated in                                           
the North-South direction (the remains of the main cluster                                             
gas peak mixed with gas stripped from the bullet) was                                                  
modeled as a pancake-like body seen from the edge, with the axes in the sky plane allowed to vary.                                               
The density profiles                                             
of these components were chosen to be broken power laws,                                               
such that each was effectively truncated at some finite                                                
radius.  The pre-shock gas was described with a beta model centered                                             
on the mass peak of the main cluster.  The emission       
to the east of the main cluster peak was fit with a second                                             
beta model with an independent scale factor along the                                                  
direction of the merger axis, so that this component was                                               
allowed to stretch or compress along the                                                   
east-west axis.  Its centroid was also allowed to vary.  The                                           
models were blended together to generate a continuous                                           
mass distribution that yielded a smooth image for fitting to                                           
the X-ray data.                                                                                        
                                                                                                       
An X-ray flux map was derived from the resulting 3D gas                                                
model by assuming that the X-ray flux was proportional to                                              
$\rho_{gas}^2$ times a factor weakly dependent on                                                      
temperature, for which we used the X-ray derived temperature                                           
map\cite{MA02.1}. This factor varied by less                                              
than 5\%.  In order to fit to the data, the X-ray image was                                            
adaptively binned such that each pixel contained at least 50                                           
net counts, so that there were fewer and smaller pixels in                                             
regions with higher signal-to-noise (the same binning was                                              
applied to the model image during fitting).  Due to the                                                
large number of parameters in our model, we used a simulated                                           
annealing technique to minimize the $\chi^2$.  For each set of          
parameters, a new 3D model was generated, and from the model                                           
a new flux map that could be compared with the data.                                                   
Finally, the overall normalization of the mass model was                                               
determined by equating the total X-ray flux to the observed                                            
value.                                                                                                 
                                                                                                       
By freezing                                            
various model parameters and refitting, we conclude that the                                           
projected gas mass measurements in the central regions,                                                
including the bullet area, are accurate to within about                                                
10\%.  Results outside the central 2.5 to 3 arcmin may be                                              
less accurate. Relevant to the problem at hand, there is a                                             
theoretical possibility that the hot gas is very clumpy on                                             
linear scales smaller than the resolution of the image,                                                
which would enhance its X-ray emission and lead to an                                                  
overestimate of the gas mass. However, such clumpiness has                                             
not been observed in clusters, including the nearby,                                                   
better-resolved ones, and so is extremely unlikely to have                                             
any significant effect for our gas mass estimate.         

\section{Weak Gravitational Lensing}

The goal of weak lensing is to obtain a map of the matter surface density by measuring the distortion of images of background galaxies caused by the deflection of light as it passes the cluster.  This deflection stretches the image of the galaxy preferentially in the direction perpendicular to that of the cluster's center of mass.  The resulting change in the ellipticity of the image is typically small compared to that intrinsic to the galaxy; therefore the distortion is only measurable statistically with large numbers of background galaxies.  We measure the ellipticities of the galaxies by measuring the second moments of the surface brightness, and correcting for smearing by the point spread function in the image.  The corrected ellipticities are a direct, but noisy, measurement of the reduced shear $g = \gamma/(1-\kappa)$.  The shear $\gamma$ is the anisotropic stretching of the galaxy image while
the convergence $\kappa$ is the shape-independent change in the size of the image.

Both $\gamma$ and $\kappa$ are second derivatives of the surface potential, which for Newtonian gravity is the gravitational potential integrated along the line of sight and $\kappa$ becomes the surface density of the system divided by a constant which depends on the angular distances between the observer, the lensing cluster, and the background galaxy.  Several methods are known which convert a reduced shear field into a $\kappa$ distribution.  For alternative gravities, however, $\kappa$ is no longer a measurement of the surface density, but is a non-local function whose overall level is still tied to the amount of mass.  For complicated system geometries, such as a merging cluster, the multiple peaks can deflect, suppress, or enhance some of the peaks.

In weak lensing, the primary source of noise is the intrinsic ellipticity of the background galaxies, and therefore to maximize the signal--to--noise of a measurement one needs to obtain very deep optical images to get the largest number density of background galaxies possible.  However, because weak lensing can only measure the change in $\kappa$ relative to the outer edge of the data, one also needs to obtain shear measurements over a region larger than the expected size of the cluster.  We therefore obtained three sets of optical images: a $34^\prime \times 34^\prime$ field--of--view set using the ESO/MPG Wide Field Imager in which we could measure shears for 15 galaxies per sq.~arcmin., a  $8^\prime$ radius field--of--view set using IMACS on Magellan which had 35 galaxies per sq.~arcmin., and two pointings of ACS on HST which gave us a $6^\prime \times 3.5^\prime$ area with 73  galaxies per sq.~arcmin.  We measured the shapes of the galaxies independently on each image set, then averaged the measurements for the galaxies in common, appropriately increasing the relative weight of the measurement.

The resulting catalog was converted into a measurement of the 2-D $\kappa$ distribution by combining various derivatives of the reduced shear to obtain a measurement of gradient of $\ln (1-\kappa)$, then integrating across the field.  The distribution is shown in the figure in contours, and as can be seen the peaks in $\kappa$ are in good spatial agreement with the galaxies in the system and not with the plasma.  We measured errors on this separation by creating bootstrap samplings of the reduced shear catalog, doing the same $\kappa$ reconstruction on the new catalogs, and measuring the centroid of the two peaks.  We did this 200,000 times and showed that the $4\sigma$ error contours reached only halfway to the gas peaks, so noise can not be causing the offset at a $\sim 8\sigma$ level (provided the contour spacing does not increase for higher significance levels).  Both the X-ray and optical images are deep enough to detect any other structures along the line of sight which could be contributing to the $\kappa$ distribution, and we do not detect any such structures.  The $\kappa$ map could still be affected by long, low-density filaments which are aligned along the line of sight, but the geometric probability of two such filaments running perfectly along the line of sight is very low, $\sim 10^{-6}$.

\section{Discussion}
From the figure one can see that the $\kappa$ peaks are in agreement with the galaxy positions and offset from the gas.  One would expect that this indicates that dark matter much be present regardless of the gravitational force law, but in some alternative gravity models, the multiple peaks can alter the lensing surface potential so that the strength of the peaks is no longer directly related to the matter density in them.  As such, all of the alternative gravity models have to be tested individually against the observations.  All of the models, however, have to pass three tests to be able to explain the system.

First, the two observed lensing peaks have $\kappa$ to optical light and X-ray plasma mass ratios which are consistent to better than a factor of 2 with those observed in other clusters which have the galaxies and plasma spatially coincident.  This means that the model has to allow for the $\sim 10\%$ of baryons which remain in a distribution whose peak is coincident with the lensing peak to create a weak lensing signal that is similar to a normal system which has 10 times the baryonic mass in the same peak.  Further, one can see that the contours are extended over the position of the gas, and measuring the resulting skewness of the peaks indicates that the gas contributes only $\sim 12\%$ of the lensing signal.

Second, for this merging cluster system, the lensing and plasma peaks are not in a symmetric arrangement.  The plasma peaks are located above the line connecting the lensing peaks, and one of the merging cluster component is several times more massive than the other.  This asymmetry will create problems for any gravity model which tries to use the two plasma peaks to deflect the lensing potentials off of the plasma.  Also note that the measurements of the X-ray plasma mass and the baryonic mass of the galaxies are independent of assumptions regarding the nature of the gravitational force law, so cannot be changed to allow for a better fit to the $\kappa$ distribution.

Third, weak lensing has been used to observe a large mass range of systems, from galaxies to groups to rich clusters, and the overall lensing strength of the system has been found to have a direct, almost linear relation to the total baryonic mass of the system.  Therefore the lensing models with alternative gravity for simple systems must scale close to linearly with the baryonic mass.

It is difficult to see how any of the proposed alternative gravity models can pass these tests without having a large majority ($>70\%$) of the mass of the system in dark matter.  Indeed, in a recent paper, a test of the TeVeS alternative gravity theory has been performed and has shown that it can only reproduce the lensing observations if the system has a dark matter component which is at least 2.4 times more massive the the baryonic mass \cite{AN06.1}.

\end{document}